\pdfoutput=1
\RequirePackage{ifpdf}
\ifpdf 
\documentclass[pdftex]{sigma}
\else
\documentclass{sigma}
\fi

\numberwithin{equation}{section}

\newcommand{\bla}{\bar u}
\newcommand{\bmu}{\bar v}
\newcommand{\blac}{\bar{u}^{\scriptscriptstyle C}}
\newcommand{\blab}{\bar{u}^{\scriptscriptstyle B}}
\newcommand{\blacb}{\bar{u}^{\scriptscriptstyle C,B}}
\newcommand{\bmuc}{\bar{v}^{\scriptscriptstyle C}}
\newcommand{\bmub}{\bar{v}^{\scriptscriptstyle B}}
\newcommand{\bmucb}{\bar{v}^{\scriptscriptstyle C,B}}

\newcommand{\so}{\scriptscriptstyle \rm I}
\newcommand{\st}{\scriptscriptstyle \rm I\hspace{-1pt}I}

\newtheorem{thm}{Theorem}[section]
\newtheorem{lem}[thm]{Lemma}
\newtheorem{Conj}[thm]{Conjecture}
{ \theoremstyle{definition}
\newtheorem{rmk}[thm]{Remark}
}

\begin{document}

\newcommand{\arXivNumber}{1502.01966}

\allowdisplaybreaks

\renewcommand{\PaperNumber}{064}

\FirstPageHeading

\ShortArticleName{${\rm GL}(3)$-Based Quantum Integrable Composite Models. II.~Form Factors of Local Operators}

\ArticleName{$\boldsymbol{{\rm GL}(3)}$-Based Quantum Integrable Composite Models.\\
II.~Form Factors of Local Operators}

\Author{Stanislav~PAKULIAK~${}^{abc}$, Eric~RAGOUCY~${}^d$ and Nikita~A.~SLAVNOV~${}^e$}

\AuthorNameForHeading{S.~Pakuliak, E.~Ragoucy and N.A.~Slavnov}

\Address{$^a$~Institute of Theoretical \& Experimental Physics, 117259 Moscow, Russia}

\Address{$^b$~Laboratory of Theoretical Physics, JINR, 141980 Dubna, Moscow reg., Russia}
\EmailD{\href{mailto:pakuliak@jinr.ru}{pakuliak@jinr.ru}}

\Address{$^c$~Moscow Institute of Physics and Technology, 141700, Dolgoprudny, Moscow reg., Russia}

\Address{$^d$~Laboratoire de Physique Th\'eorique LAPTH,
CNRS and Universit\'e de Savoie, \\
\hphantom{$^d$}~BP 110, 74941 Annecy-le-Vieux Cedex, France}
\EmailD{\href{mailto:eric.ragoucy@lapth.cnrs.fr}{eric.ragoucy@lapth.cnrs.fr}}

\Address{${}^e$ Steklov Mathematical Institute, Moscow, Russia}
\EmailD{\href{mailto:nslavnov@mi.ras.ru}{nslavnov@mi.ras.ru}}

\ArticleDates{Received February 18, 2015, in f\/inal form July 22, 2015; Published online July 31, 2015}

\Abstract{We study  integrable models solvable by the nested algebraic Bethe ansatz and possessing the
${\rm GL}(3)$-invariant $R$-matrix. We consider a composite model where the total monodromy matrix of the model is presented as
a product of two partial monodromy matrices. Assuming that the last ones can be expanded
into series with respect to the inverse spectral parameter we calculate matrix elements of the local operators in the basis
of the transfer matrix eigenstates. We obtain determinant representations for these matrix elements.
Thus, we solve the inverse scattering problem in a weak sense.}


\Classification{17B37; 81R50}

\section{Introduction}

The algebraic Bethe ansatz was found to be a powerful method for describing the spectrum of various quantum  integrable models \cite{FadLH96,KulR83, FadST79,FadT79}. In  this approach, quantum Hamiltonians and all other integrals of motion are
generated by a transfer matrix. The eigenstates of the latest can be found in a systematic way, leading to a set of equations determining
the spectrum (Bethe equations).

Despite the signif\/icant progress of the algebraic Bethe ansatz
in  calculating the spectrum, the application of this technique to the problem of
calculating correlation functions for a long time led to much more limited results. It is worth
mentioning the papers \cite{IzeK84,IzeKR87,Kor84}, where series representations for correlations
of the model of one-dimensional bosons were obtained. Later, Fredholm determinant representations for correlation functions
of this model were derived by the method of dual f\/ields (see \cite{BogIK93L} and references therein).

Solution of the quantum inverse scattering problem \cite{KitMT99,MaiT00} has opened up new opportunities of
the algebraic Bethe ansatz. Using this result correlation functions of the $XXZ$ spin chain were studied in
series of works \cite{GohKS04,GohKS05,KitKMST09a, KitMST02,KitMST05}. The explicit formulas for the local operators
provided by the inverse scattering problem also had played an important role in  calculating their form factors
\cite{CauM05,KitKMST09b,KitKMST11}. It is worth mentioning, however, that the results of \cite{KitMT99,MaiT00} essentially
were based on the fact that the monodromy matrix of the model could be constructed from the $R$-matrix. This is true
for various spin chains, but not in general. In the present paper we use the approach of~\cite{IzeK84} for calculating form factors of local operators in
quantum ${\rm GL}(3)$-invariant models. Let us brief\/ly describe the main idea of this method.

The key equation of the quantum inverse scattering method is the $RTT$-relation \cite{FadST79,FadT79}
\begin{gather}\label{RTT}
R_{12}(u,v)T_1(u)T_2(v)=T_2(v)T_1(u)R_{12}(u,v).
\end{gather}
Here $T(u)$ is the monodromy matrix, $R(u,v)$ is the $R$-matrix. In  ${\rm GL}(3)$-invariant models the $R$-matrix
acts in the tensor product of two auxiliary spaces $V_1\otimes V_2$ ($V_k\sim\mathbb{C}^3$, $k=1,2$) and has the form
 \begin{gather}\label{R-mat}
 R(u,v)=\mathbf{I}+g(u,v)\mathbf{P},\qquad g(u,v)=\frac{c}{u-v}.
 \end{gather}
Here $\mathbf{I}$ is the identity matrix in $V_1\otimes V_2$, $\mathbf{P}$ is the permutation matrix
that exchanges~$V_1$ and~$V_2$, and $c$ is a constant. The monodromy matrix $T(u)$ acts in $\mathbb{C}^3\otimes\mathcal{H}$, where
$\mathcal{H}$ is the Hilbert space of the Hamiltonian of the model under consideration. Equation~\eqref{RTT} holds in the tensor product $V_1\otimes V_2\otimes\mathcal{H}$, and the  matrices $T_k(w)$ act non-trivially in
$V_k\otimes \mathcal{H}$.

The monodromy matrix $T(u)$ of a lattice quantum model is equal to the product of local $L$-operators
\begin{gather}\label{mat-T}
T(u)=L_M(u)\cdots L_1(u),
\end{gather}
where $M$ is the number of lattice sites, and every $L$-operator satisf\/ies the $RTT$-relation with the $R$-matrix~\eqref{R-mat}. Continuous quantum models appear in the limit $M\to\infty$. Let us f\/ix some
site~$m$ ($1\le m<M$) and def\/ine two partial monodromy matrices $T^{(1)}(u)$  and $T^{(2)}(u)$ as
\begin{gather}\label{mat-T12}
T^{(1)}(u)=L_m(u)\cdots L_1(u),\qquad T^{(2)}(u)=L_M(u)\cdots L_{m+1}(u).
\end{gather}
Then obviously
\begin{gather}\label{T-TT}
T(u)=T^{(2)}(u)T^{(1)}(u).
\end{gather}
We call such model {\it composite generalized model}\footnote{The authors of \cite{IzeK84} used the terminology {\it two-site model}.
We think that this terminology becomes misleading in the case of spin chains. The terminology {\it two-component model} used
in \cite{Sla07} also becomes misleading in the case of multi-component Bose or Fermi gases.  We choose to use the terminology introduced in
\cite{PakRS15b}.} \cite{IzeK84}.

We assume that local $L$-operators in \eqref{mat-T} depend on the spectral parameter $u$ as follows\footnote{See \cite{KulRes82} for concrete examples of $L$-operators satisfying the $RTT$-relation with the ${\rm GL}(3)$-invariant $R$-matrix.}:
\begin{gather*}
L_n(u)=\mathbf{1}+\frac cu L_n[0] +o\big(u^{-1}\big),\qquad u\to\infty.
\end{gather*}
Here $\mathbf{1}$ is the identity operator in~$\mathbb{C}^3\otimes\mathcal{H}$, and the matrix elements
of~$L_n[0]$ are local operators of the model. Then it is easy to see that
both partial monodromy matrices~$T^{(l)}(u)$ have the standard expansion over $c/u$:
\begin{gather}\label{zero-modes}
T^{(l)}(u)=\mathbf{1}+ \sum_{n=0}^\infty T^{(l)}[n] \left(\frac cu\right)^{n+1},\qquad l=1,2,
\end{gather}
where the partial zero mode $T^{(1)}[0]$ is equal to
\begin{gather}\label{part-zm}
T^{(1)}[0]=\sum_{n=1}^m L_n[0].
\end{gather}

In this paper we develop a method of calculating form factors of matrix elements~$T^{(1)}_{ij}[0]$. We  reduce them to the form factors of the monodromy matrix entries $T_{ij}(z)$ studied in our previous publications. In this way we obtain determinant
representations for the form factors of~$T^{(1)}_{ij}[0]$   and extract explicitly their dependence on the  lattice site number~$m$. Then, taking
the lattice derivative of the results obtained, we f\/ind determinant representations for the form factors of the
local operators~$(L_m[0])_{ij}$.  We do not give here explicit determinant formulas, but the reader can f\/ind them in~\cite{BelPRS12b,BelPRS13a,PakRS14c,PakRS14b,PakRS15a}.

The paper is organized as follows. In Section~\ref{S-GM} we introduce basic notions of the generalized model. In
Section~\ref{S-CGM} we consider composite  generalized model.  Section~\ref{S-MR} contains the main results of this paper. There we
reduce form factors of local operators to the ones of the monodromy matrix entries
for the ${\rm GL}(3)$ case and we conjecture a form for the ${\rm GL}(N)$ case.
The following sections contain the proofs for ${\rm GL}(3)$. In  Section~\ref{S-FFDO} we consider the form factors of the diagonal partial zero modes $T^{(1)}_{ii}[0]$.  Finally, in Section~\ref{S-FFT12} we study the form factors of the partial zero modes
$T^{(1)}_{ij}[0]$ for $i\ne j$.

\section{Generalized model}\label{S-GM}

\subsection{Bethe vectors}\label{SS-BV}

In the framework of the algebraic Bethe ansatz the entries of $T(w)$ act in a Hilbert space $\mathcal{H}$ and its dual $\mathcal{H}^*$  that possess
a pseudovacuum vector $|0\rangle$ and  a dual pseudovacuum vector $\langle 0|$ respectively.
They are normalized by the condition $\langle 0|0\rangle=1$.
These vectors
are annihilated by the operators $T_{ij}(w)$, where $i>j$ for  $|0\rangle$ and $i<j$ for $\langle0|$.
At the same time both vectors are eigenvectors for the diagonal entries of the monodromy matrix
 \begin{gather*}
 T_{ii}(w)|0\rangle=\lambda_i(w)|0\rangle, \qquad   \langle0|T_{ii}(w)=\lambda_i(w)\langle0|,\qquad i=1,2,3,
 \end{gather*}
where $\lambda_i(w)$ are some scalar functions. In the framework of the generalized model, $\lambda_i(w)$ remain free functional parameters. Actually, it is always possible to normalize
the monodromy matrix $T(w)\to \lambda_2^{-1}(w)T(w)$ so as to deal only with the ratios
 \begin{gather}\label{ratios}
 r_1(w)=\frac{\lambda_1(w)}{\lambda_2(w)}, \qquad  r_3(w)=\frac{\lambda_3(w)}{\lambda_2(w)}.
 \end{gather}
Below we assume that $\lambda_2(w)=1$.

Bethe vectors are certain polynomials in the operators~$T_{ij}(u)$ with $i<j$ acting on the pseudovacuum vector
$|0\rangle$ \cite{BelPRS12c,KhoPak08,KhoPakT07,KulR83,TarVar93}. In the ${\rm GL}(3)$-invariant models they depend on two sets of variables called Bethe parameters.
We denote the Bethe vectors~$\mathbb{B}_{a,b}(\bla;\bmu)$. Here the Bethe parameters are $\bla=\{u_1,\dots,u_a\}$ and
$\bmu=\{v_1,\dots,v_b\}$. The subscripts $a$ and $b$ ($a,b=0,1,\dots$) respectively denote the cardinalities of the sets~$\bla$ and~$\bmu$.

Similarly we can construct dual Bethe vectors in the dual space as polynomials in the ope\-ra\-tors~$T_{ij}(u)$ with $i>j$ acting on the
dual pseudovacuum vector~$\langle0|$. We denote them~$\mathbb{C}_{a,b}(\bla;\bmu)$ with the same meaning of the arguments and
subscripts.

\subsection{Notation}\label{SS-Not}

Besides the function $g(u,v)$ we also introduce a function $f(u,v)$
\begin{gather}\label{univ-not}
 f(u,v)=\frac{u-v+c}{u-v}.
\end{gather}

We denote sets of variables by bar:  $\bla$, $\bmu$ etc. If necessary, the cardinalities of the sets are
given in special comments.
Individual elements of the sets are denoted by subscripts: $w_j$, $u_k$ etc.
We say that $\bar x=\bar x'$,
if $\#\bar x=\#\bar x'$ and $x_i=x'_i$ (up to a permutation) for $i=1,\dots,\#\bar x$. We say that $\bar x\ne \bar x'$ otherwise.

Below we consider partitions of sets into subsets.  The notation $\bar u\Rightarrow\{\bla_{\so},\bla_{\st}\}$ means that the set $\bla$
is divided into two disjoint subsets. As a rule, we use roman numbers for subscripts of subsets: $\bla_{\so}$, $\bmu_{\rm ii}$ etc.
However, if we deal with a big quantity of subsets, then we use standard arabic numbers for their notation. In such cases we give special comments  to avoid ambiguities.

Similarly to our previous papers (see, for instance, \cite{PakRS15b}) we use a shorthand notation for products of  some functions.
Namely, if the functions $r_k$ \eqref{ratios} or the function $f$ \eqref{univ-not} depend
on sets of variables, this means that one should take the product over the corresponding set.
For example,
 \begin{gather}\label{SH-prod}
 r_1(\bla)=\prod_{u_k\in\bla} r_1(u_k),\qquad
 f(z, \bar w)= \prod_{w_j\in\bar w} f(z, w_j),\qquad
 f(\bla,\bmu)=\prod_{u_j\in\bla}\prod_{v_k\in\bmu} f(u_j,v_k).
 \end{gather}
By def\/inition any product with respect to the empty set is equal to $1$. If we have a double product, then it is also equal to $1$ if at least
one of sets is empty.

In Section~\ref{S-CGM} we shall introduce several new scalar functions and will extend
the convention~\eqref{SH-prod} to their products.

\subsection{On-shell Bethe vectors}\label{SS-OBV}

In the algebraic Bethe
ansatz the role of a quantum Hamiltonian is played by  the transfer matrix. It is the trace in the auxiliary space of the monodromy matrix: $\operatorname{tr} T(u)$.
The eigenstates of the transfer matrix are called {\it on-shell} Bethe vectors. The eigenstates of the transfer matrix in the dual space are called
dual on-shell Bethe vectors\footnote{%
For simplicity here and below we do not distinguish between vectors and dual vectors, because their properties are completely analogous to each other.}.
 We will denote usual and dual Bethe vectors as $\mathbb{B}_{a,b}(\bla,\bmu)$ and $\mathbb{C}_{a,b}(\bla,\bmu)$ respectively. Dif\/ferent equivalent
formulas for these Bethe vectors were presented in the paper \cite{BelPRS12c}. In what follows  we will not use these explicit expressions. Instead, we will
use formulas which relate Bethe vectors of the composite model with the ones for the components of the model (see formulas \eqref{BV-BV} and \eqref{CV-CV} which are
proved in the f\/irst part of this paper \cite{PakRS15b}) and the action of the
monodromy matrix elements onto Bethe vectors obtained in \cite{BelPRS12c}.

A (dual) Bethe vector becomes on-shell, if the Bethe parameters satisfy the system of Bethe equations.
We give this system in a slightly unusual form
\begin{gather}\label{AEigenS-1}
r_1(\bla_{\so})=\frac{f(\bla_{\so},\bla_{\st})}{f(\bla_{\st},\bla_{\so})}f(\bmu,\bla_{\so}),\qquad
r_3(\bmu_{\so})=\frac{f(\bmu_{\st},\bmu_{\so})}{f(\bmu_{\so},\bmu_{\st})}f(\bmu_{\so},\bla).
\end{gather}
These equations should hold for arbitrary partitions of the sets $\bla$ and $\bmu$ into subsets
$\{\bla_{\so}, \bla_{\st}\}$ and $\{\bmu_{\so}, \bmu_{\st}\}$ respectively. Obviously, it is enough
to demand that the system  \eqref{AEigenS-1} is valid for the
particular case, when the sets $\bla_{\so}$ and $\bmu_{\so}$ consist of only one element. Then it  turns into the standard system of Bethe equations.

If the sets $\bla$ and $\bmu$ satisfy \eqref{AEigenS-1}, then
\begin{gather*}
\operatorname{tr} T(w) \mathbb{B}_{a,b}(\bla,\bmu)= \tau(w|\bla,\bmu)\mathbb{B}_{a,b}(\bla,\bmu),\qquad
 \mathbb{C}_{a,b}(\bla,\bmu)\operatorname{tr} T(w)= \tau(w|\bla,\bmu)\mathbb{C}_{a,b}(\bla,\bmu),
 \end{gather*}
with
\begin{gather*}
\tau(w|\bla,\bmu)=r_1(w)f(\bla,w)+f(w,\bla)f(\bmu,w)+r_3(w)f(w,\bmu).
\end{gather*}

Besides usual on-shell Bethe vectors it is also convenient to consider {\it twisted on-shell} Bethe vectors (see, e.g., \cite{BelPRS12b}). They are eigenstates
of a twisted transfer matrix, that in its turn, is the  trace in the auxiliary space of the twisted monodromy matrix $T_{\bar\kappa}(u)$. The last one
is def\/ined as $T_{\bar\kappa}(u)=\hat\kappa T(u)$, where
 $\hat\kappa=\operatorname{diag}(\kappa_1,\kappa_2, \kappa_3)$. The matrix elements~$\kappa_i$ ($i=1,2,3$) are called twist parameters.
A (dual) Bethe vector becomes twisted on-shell vector, if the Bethe parameters satisfy the system of twisted Bethe equations:
\begin{gather}\label{ATEigenS-1}
r_1(\bla_{\so})=\left(\frac{\kappa_2}{\kappa_1}\right)^{k_{\so}}\frac{f(\bla_{\so},\bla_{\st})}{f(\bla_{\st},\bla_{\so})}f(\bmu,\bla_{\so}),\qquad
r_3(\bmu_{\so})=\left(\frac{\kappa_2}{\kappa_3}\right)^{n_{\so}}\frac{f(\bmu_{\st},\bmu_{\so})}{f(\bmu_{\so},\bmu_{\st})}f(\bmu_{\so},\bla),
\end{gather}
where $k_{\so}=\#\bla_{\so}$ and $n_{\so}=\#\bmu_{\so}$. These equations also should hold for arbitrary partitions of the sets $\bla$ and $\bmu$ into subsets.

\subsection{Description of scalar products}

Recall a formula for the scalar product of generic Bethe vectors \cite{Res86}:
 \begin{gather}
 \mathcal{S}_{a,b}\equiv \mathbb{C}_{a,b}(\blac;\bmuc)\mathbb{B}_{a,b}(\blab;\bmub)=\sum r_1(\blab_{\so})r_1(\blac_{\st})
 r_3(\bmuc_{\st})r_3(\bmub_{\so})
  f(\blac_{\so},\blac_{\st})  f(\blab_{\st},\blab_{\so})    \label{Resh-SP} \\
\hphantom{\mathcal{S}_{a,b}\equiv}{}
 \times f(\bmuc_{\st},\bmuc_{\so})   f(\bmub_{\so},\bmub_{\st})\frac{f(\bmuc_{\so},\blac_{\so})f(\bmub_{\st},\blab_{\st})}{f(\bmuc,\blac)f(\bmub,\blab)}
  Z_{a_{\st},b_{\so}}(\blac_{\st};\blab_{\st}|\bmuc_{\so};\bmub_{\so})
 Z_{a_{\so},b_{\st}}(\blab_{\so};\blac_{\so}|\bmub_{\st};\bmuc_{\st}).\nonumber
 \end{gather}
Here all the Bethe parameters are generic complex numbers and the sum is taken over the partitions of the sets $\blac$, $\blab$, $\bmuc$, and $\bmub$
 \begin{alignat*}{3}
&  \blac\Rightarrow\{\blac_{\so}, \blac_{\st}\},  \qquad &&  \bmuc\Rightarrow\{\bmuc_{\so}, \bmuc_{\st}\},& \\
& \blab\Rightarrow\{\blab_{\so}, \blab_{\st}\},  \qquad  && \bmub\Rightarrow\{\bmub_{\so}, \bmub_{\st}\} .&
 \end{alignat*}

The partitions are independent except that $\#\blab_{\so}=\#\blac_{\so}=a_{\so}$ with $a_{\so}=0,\dots,a$, and $\#\bmub_{\so}=\#\bmuc_{\so}=b_{\so}$
with $b_{\so}=0,\dots,b$. From this we f\/ind $\#\blab_{\st}=\#\blac_{\st}=a_{\st}=a-a_{\so}$  and
$\#\bmub_{\st}=\#\bmuc_{\st}=b_{\st}=b-b_{\so}$. The functions  $Z_{a_{\st},b_{\so}}$  and $Z_{a_{\so},b_{\st}}$ are so-called highest coef\/f\/icients.
They are equal to a partition function of $15$-vertex model with special boundary conditions~\cite{Res86}. The reader can f\/ind their
explicit representations in~\cite{BelPRS12a, Whe12}. We do not use these explicit formulas in the present paper  except
$Z_{0,0}(\varnothing;\varnothing|\varnothing;\varnothing)=1$. This condition is needed to satisfy the normalization $\mathcal{S}_{0,0}=\langle0|0\rangle=1$.

If $\mathbb{C}_{a,b}(\blac;\bmuc)$ is a twisted on-shell vector and $\mathbb{B}_{a,b}(\blab;\bmub)$ is a usual on-shell vector,
then we can express the functions $r_1(u^{\scriptscriptstyle C,B}_j)$ and $r_3(v^{\scriptscriptstyle C,B}_j)$ in~\eqref{Resh-SP} in terms of (twisted) Bethe equa\-tions~\eqref{AEigenS-1},~\eqref{ATEigenS-1}.
We denote such  scalar product
by~$\mathcal{S}_{a,b}^{(\bar\kappa)}$. It is easy to see that
 \begin{gather}
 \mathcal{S}_{a,b}^{(\bar\kappa)}=\sum  \left(\frac{\kappa_2}{\kappa_1}\right)^{a_{\st}}
 \left(\frac{\kappa_2}{\kappa_3}\right)^{b_{\st}}
  f(\blac_{\st},\blac_{\so})  f(\blab_{\so},\blab_{\st})  f(\bmuc_{\so},\bmuc_{\st})
   f(\bmub_{\st},\bmub_{\so})
  \nonumber\\
\hphantom{\mathcal{S}_{a,b}^{(\bar\kappa)}=}{}
 \times f(\bmuc_{\st},\blac_{\st})f(\bmub_{\so},\blab_{\so})  Z_{a_{\st},b_{\so}}(\blac_{\st};\blab_{\st}|\bmuc_{\so};\bmub_{\so})
 Z_{a_{\so},b_{\st}}(\blab_{\so};\blac_{\so}|\bmub_{\st};\bmuc_{\st}).\label{Resh-SP-eig}
 \end{gather}

\begin{rmk}
\label{Rem1}
One should be careful using Bethe equations in the sums over partitions, because there might be problems if some parameters from
the sets~$\blac$,~$\bmuc$ coincide with the ones from~$\blab$,~$\bmub$ (i.e., $\blac\cap\blab\ne\varnothing$ or/and $\bmuc\cap\bmub\ne\varnothing$). The matter is that after imposing the Bethe equations we cannot consider the
limit where one solution of Bethe equations goes to another solution.
Instead one should f\/irst take the limit in~\eqref{Resh-SP}  (what leads to the appearance of the derivatives~$r'_1(u)$ and~$r'_3(v)$) and only then impose Bethe
equations. However, if we consider the scalar product of twisted and usual on-shell vectors, then we can use the twisted and
usual Bethe equations from the very beginning. The matter is that in this case the parameters~$\blac$ and~$\bmuc$ are
functions of~$\bar\kappa=\{\kappa_1,\kappa_2,\kappa_3\}$: $\blac=\blac(\bar\kappa)$ and $\bmuc=\bmuc(\bar\kappa)$. Therefore we always can take~$\bar\kappa$ such that
$\blac\cap\blab=\varnothing$ and $\bmuc\cap\bmub=\varnothing$. Then, if necessary, we can consider the limit where some
parameters coincide. In this case we should treat the variables~$\blac$ and~$\bmuc$ as functions of~$\bar\kappa$.
\end{rmk}

The sum over partitions~\eqref{Resh-SP-eig} was studied in~\cite{BelPRS12b,BelPRS13a} for {\it arbitrary} values of the Bethe parameters. There this sum was reduced to a single determinant in the case $\kappa_1=\kappa_3$. It was proved that the determinant vanishes
at\footnote{Here and below the notation $\bar\kappa=1$ means $\kappa_i=1$, $i=1,2,3$.} $\bar\kappa=1$. Thus,  if~$\blacb\ne\varnothing$
 or~$\bmucb\ne\varnothing$, then
setting $\bar\kappa=1$ in~\eqref{Resh-SP-eig} we obtain an identity
 \begin{gather}
 0=\sum    f(\blac_{\st},\blac_{\so})  f(\blab_{\so},\blab_{\st})  f(\bmuc_{\so},\bmuc_{\st})
   f(\bmub_{\st},\bmub_{\so})f(\bmuc_{\st},\blac_{\st})f(\bmub_{\so},\blab_{\so})
  \nonumber\\
 \hphantom{0=}{}
 \times   Z_{a_{\st},b_{\so}}(\blac_{\st};\blab_{\st}|\bmuc_{\so};\bmub_{\so})
 Z_{a_{\so},b_{\st}}(\blab_{\so};\blac_{\so}|\bmub_{\st};\bmuc_{\st}).\label{Resh-SP-eig2}
 \end{gather}
In the exceptional case $\blacb=\bmucb=\varnothing$ we have
 \begin{gather}\label{Exception}
 \mathcal{S}_{0,0}^{(\bar\kappa)}\Bigr|_{\bar\kappa=1}=1.
 \end{gather}

\begin{rmk}\label{Rem2}
One should not be surprised that the r.h.s.\ of~\eqref{Resh-SP-eig2} does not give the norm of an on-shell Bethe vector in the case
$\{\blac,\bmuc\}=\{\blab,\bmub\}$. Indeed, as we explained above, in order to obtain the norm one should consider the limit
$\{\blac(\bar\kappa),\bmuc(\bar\kappa)\}\to\{\blab,\bmub\}$  at $\bar\kappa\to 1$  in~\eqref{Resh-SP-eig}. Instead we simply set
$\bar\kappa= 1$ in \eqref{Resh-SP-eig} for generic values of the Bethe parameters. In this case we obtain that the sum~\eqref{Resh-SP-eig2} vanishes~\cite{BelPRS12b}, and then this result can be continued to the point $\{\blac,\bmuc\}=\{\blab,\bmub\}$.
\end{rmk}

 Identity \eqref{Resh-SP-eig2} plays a central role in the calculation of form factors of local operators.

\subsection{Universal form factors}

Form factors of the monodromy matrix entries are def\/ined as
 \begin{gather*}
 \mathcal{F}_{a,b}^{(i,j)}(z)\equiv\mathcal{F}_{a,b}^{(i,j)}(z\,|\,\blac,\bmuc;\blab,\bmub)=
 \mathbb{C}^{a',b'}(\blac;\bmuc)T_{ij}(z)\mathbb{B}^{a,b}(\blab;\bmub),
 \end{gather*}
where both $\mathbb{C}^{a',b'}(\blac;\bmuc)$ and $\mathbb{B}^{a,b}(\blab;\bmub)$ are on-shell
Bethe vectors, and
\begin{gather*}
a'=a+\delta_{i1}-\delta_{j1},\qquad
b'=b+\delta_{j3}-\delta_{i3}.
\end{gather*}
The parameter $z$ is an arbitrary complex  number.

It was proved in \cite{PakRS15a} that if $\{\blac,\bmuc\}\ne \{\blab,\bmub\}$, then  the combination
\begin{gather*}
\mathfrak{F}_{a,b}^{(i,j)}(\blac,\bmuc;\blab,\bmub)=
\frac{\mathcal{F}_{a,b}^{(i,j)}(z\,|\,\blac,\bmuc;\blab,\bmub)}
{\tau(z\,|\,\blac,\bmuc)-\tau(z\,|\,\blab,\bmub)}
\end{gather*}
does not depend on $z$. We call $\mathfrak{F}_{a,b}^{(i,j)}(\blac,\bmuc;\blab,\bmub)$  the {\it universal form
factor} of the operator $T_{ij}(z)$. If $\blac\cap\blab=\varnothing$ and $\bmuc\cap\bmub=\varnothing$, then
the universal form factor is determined by the $R$-matrix only. It does not depend on a specif\/ic model,
in particular, on the functions~$r_1(z)$ and~$r_3(z)$.

\section{Composite  generalized model}\label{S-CGM}

Consider  a composite  generalized model def\/ined by \eqref{mat-T12}, \eqref{T-TT}.
Every $T^{(l)}(u)$ satisf\/ies $RTT$-relation \eqref{RTT} and has its own vacuum state~$|0\rangle^{(l)}$. Hereby
$|0\rangle= |0\rangle^{(1)}\otimes|0\rangle^{(2)}$.
The opera\-tors~$T_{ij}^{(2)}(u)$ and $ T_{kl}^{(1)}(v)$ commute with each other, as they act in dif\/ferent spaces.

Let
\begin{gather}\label{eigen}
T_{ii}^{(l)}(u)|0\rangle^{(l)}= \lambda_{i}^{(l)}(u)|0\rangle^{(l)}, \qquad l=1,2.
\end{gather}
We also introduce
\begin{gather*}
r_{k}^{(l)}(u)=\frac{\lambda_{k}^{(l)}(u)}{\lambda_{2}^{(l)}(u)} \qquad l=1,2, \quad k=1,3.
\end{gather*}
Obviously
\begin{gather*}
\lambda_{i}(u)=\lambda_{i}^{(1)}(u)\lambda_{i}^{(2)}(u), \qquad
r_{k}(u)=r_{k}^{(1)}(u)r_{k}^{(2)}(u).
\end{gather*}
Below we  express form factors in terms of $r_{k}^{(1)}(u)$, therefore we introduce a special notation for
these functions
\begin{gather}\label{ell}
r_{k}^{(1)}(u)=\ell_k(u),\qquad \text{and hence,}\qquad r_{k}^{(2)}(u)=\frac{r_k(u)}{\ell_k(u)}, \qquad k=1,3.
\end{gather}
Observe that \eqref{zero-modes} implies
\begin{gather}\label{zero-modesl}
\ell_k(u)=1+ \ell_k[0] \frac cu +o\big(u^{-1}\big),
\end{gather}
and due to \eqref{T-TT}
\begin{gather*}
r_k(u)=1+ r_k[0] \frac cu +o\big(u^{-1}\big).
\end{gather*}
We extend convention~\eqref{SH-prod} to the products of the functions~$r_{k}^{(l)}(u)$ and~$\ell_k(u)$. Namely, whenever these
functions depend on sets of variables (for instance $r_{k}^{(l)}(\bla)$ or $\ell_k(\bmu_{\st})$) this means the product over the
corresponding set.

\subsection{Bethe vectors and partial Bethe vectors}

We can introduce partial Bethe vectors $\mathbb{B}^{(l)}_{a,b}(\bla;\bmu)$ for both partial monodromy matrices $T^{(l)}(u)$.
Then Bethe vectors of the total monodromy matrix can be expressed in terms of the partial Bethe vectors as follows~\cite{BeKhP07,PakRS15b}:
\begin{gather}\label{BV-BV}
\mathbb{B}_{a,b}(\bla;\bmu)=\sum r_{1}^{(2)}(\bla_{\so})r_{3}^{(1)}(\bmu_{\st}) \frac{f(\bla_{\st},\bla_{\so})f(\bmu_{\st},\bmu_{\so})}{f(\bmu_{\st},\bla_{\so})}
\mathbb{B}_{a_{\so},b_{\so}}^{(1)}(\bla_{\so};\bmu_{\so}) \mathbb{B}_{a_{\st},b_{\st}}^{(2)}(\bla_{\st};\bmu_{\st}).
\end{gather}
The sum is taken over partitions $\bla\Rightarrow\{\bla_{\so},\bla_{\st}\}$ and $\bmu\Rightarrow\{\bmu_{\so},\bmu_{\st}\}$. The cardinalities of the subsets are given by the subscripts of the Bethe vectors.

We can present the product of functions $r_{1}^{(2)}(\bla_{\so})$ as
$r_{1}^{(2)}(\bla_{\so})=r_1(\bla_{\so})\ell^{-1}_1(\bla_{\so})$, see \eqref{ell}. Moreover, if we deal with an on-shell Bethe vector, we can express $r_1(\bla_{\so})$ in terms of the function $f$, thanks to the Bethe equations
\eqref{AEigenS-1}. Then we obtain
\begin{gather}\label{BV-BV-1}
\mathbb{B}_{a,b}(\bla;\bmu)=\sum \frac{\ell_{3}(\bmu_{\st})}{\ell_{1}(\bla_{\so})} f(\bla_{\so},\bla_{\st})f(\bmu_{\st},\bmu_{\so})f(\bmu_{\so},\bla_{\so})
\mathbb{B}_{a_{\so},b_{\so}}^{(1)}(\bla_{\so};\bmu_{\so}) \mathbb{B}_{a_{\st},b_{\st}}^{(2)}(\bla_{\st};\bmu_{\st}).
\end{gather}

Similarly, dual Bethe vectors can be expressed in terms of partial dual Bethe vectors
\begin{gather}\label{CV-CV}
\mathbb{C}_{a,b}(\bla;\bmu)=\sum r_{1}^{(1)}(\bla_{\st}) r_{3}^{(2)}(\bmu_{\so})\frac{f(\bla_{\so},\bla_{\st})f(\bmu_{\so},\bmu_{\st})}{f(\bmu_{\so},\bla_{\st})}
\mathbb{C}_{a_{\so},b_{\so}}^{(1)}(\bla_{\so};\bmu_{\so}) \mathbb{C}_{a_{\st},b_{\st}}^{(2)}(\bla_{\st};\bmu_{\st}),
\end{gather}
where the sum is taken again over partitions $\bla\Rightarrow\{\bla_{\so},\bla_{\st}\}$ and $\bmu\Rightarrow\{\bmu_{\so},\bmu_{\st}\}$.

If $\mathbb{C}_{a,b}(\bla;\bmu)$ is a twisted on-shell Bethe vector, then we can  use again \eqref{ell} as
$r_{3}^{(2)}(\bmu_{\so})=r_3(\bmu_{\so})\ell^{-1}_3(\bmu_{\so})$  and express $r_3(\bmu_{\so})$ through the twisted
Bethe equations \eqref{ATEigenS-1}. We get
\begin{gather}\label{CV-CV-eig}
\mathbb{C}^{(\bar\kappa)}_{a,b}   (\bla;\bmu)=\sum \left(\frac{\kappa_2}{\kappa_3}\right)^{b_{\so}}\frac{\ell_{1}(\bla_{\st})} {\ell_{3}(\bmu_{\so})}f(\bla_{\so},\bla_{\st})f(\bmu_{\st},\bmu_{\so})f(\bmu_{\so},\bla_{\so})
\mathbb{C}_{a_{\so},b_{\so}}^{(1)}(\bla_{\so};\bmu_{\so}) \mathbb{C}_{a_{\st},b_{\st}}^{(2)}(\bla_{\st};\bmu_{\st}).
\end{gather}
Here we have added the superscript $(\bar\kappa)$ to the vector $\mathbb{C}^{(\bar\kappa)}_{a,b}(\bla;\bmu)$ in order to
stress that it is a~twisted dual on-shell Bethe vector.

\subsection{The action of total and partial zero modes}

The action of the operators $T^{(l)}_{ij}(z)$ on the corresponding partial Bethe vectors $\mathbb{B}^{(l)}_{a,b}(\bla;\bmu)$ is the same
as the action of total $T_{ij}(z)$ on the total Bethe vectors $\mathbb{B}_{a,b}(\bla;\bmu)$ \cite{BelPRS12c}. One should only replace in the
formulas the functions $r_k(z)$ by their partial analogs $r^{(l)}_k(z)$. The same replacement should be done in the action
of the partial zero modes on the partial Bethe vectors \cite{PakRS15a}. In this section we give some of those actions used below.

The action of the total zero modes  $T_{ij}[0]$ (with $i<j$) on the
total Bethe vectors $\mathbb{B}_{a,b}$ can be easily extracted from the formulas given in Appendix~A of the f\/irst part of this paper~\cite{PakRS15b}
using  expansion of the monodromy matrix elements \eqref{zero-modes}. They are
\begin{gather}
T_{13}[0]\mathbb{B}_{a,b}(\bla;\bmu)=\lim_{w\to\infty}\tfrac wc \mathbb{B}_{a+1,b+1}(\{w,\bla\};\{w,\bmu\}),\nonumber \\
T_{12}[0]\mathbb{B}_{a,b}(\bla;\bmu)=\lim_{w\to\infty} \tfrac wc \mathbb{B}_{a+1,b}(\{w,\bla\};\bmu),\nonumber \\
T_{23}[0]\mathbb{B}_{a,b}(\bla;\bmu)=\lim_{w\to\infty} \tfrac wc \mathbb{B}_{a,b+1}(\bla;\{w,\bmu\}).\label{actBVa0}
\end{gather}
The right action of the operators $T_{ji}[0]$ with $i<j$ on dual Bethe vectors is quite analogous. One should replace
in \eqref{actBVa0} $T_{ij}[0]$ by $T_{ji}[0]$ and $\mathbb{B}_{a,b}(\bla;\bmu)$ by $\mathbb{C}_{a,b}(\bla;\bmu)$.

The action of the partial zero modes  $T^{(1)}_{ij}[0]$ (with $i<j$) on the
partial Bethe vectors $\mathbb{B}^{(1)}_{a,b}$ is  similar to~\eqref{actBVa0}:
\begin{gather}
T^{(1)}_{13}[0]\mathbb{B}^{(1)}_{a,b}(\bla;\bmu)=\lim_{w\to\infty}\tfrac wc \mathbb{B}^{(1)}_{a+1,b+1}(\{w,\bla\};\{w,\bmu\}), \nonumber\\
T^{(1)}_{12}[0]\mathbb{B}^{(1)}_{a,b}(\bla;\bmu)=\lim_{w\to\infty} \tfrac wc \mathbb{B}^{(1)}_{a+1,b}(\{w,\bla\};\bmu), \nonumber\\
T^{(1)}_{23}[0]\mathbb{B}^{(1)}_{a,b}(\bla;\bmu)=\lim_{w\to\infty} \tfrac wc \mathbb{B}^{(1)}_{a,b+1}(\bla;\{w,\bmu\}).\label{actBVa}
\end{gather}

The action of the  partial zero modes $T^{(1)}_{ii}[0]$ has the following form
\begin{gather}
T^{(1)}_{11}[0]\mathbb{B}^{(1)}_{a,b}(\bla;\bmu)=(\ell_1[0]-a)\mathbb{B}^{(1)}_{a,b}(\bla;\bmu), \nonumber\\
T^{(1)}_{22}[0]\mathbb{B}^{(1)}_{a,b}(\bla;\bmu)=(a-b)\mathbb{B}^{(1)}_{a,b}(\bla;\bmu), \nonumber\\
T^{(1)}_{33}[0]\mathbb{B}^{(1)}_{a,b}(\bla;\bmu)=(\ell_3[0]+b) \mathbb{B}^{(1)}_{a,b}(\bla;\bmu),\label{actBVd}
\end{gather}
where $\ell_k[0]$ are determined by \eqref{zero-modesl}.
In all the formulas above Bethe vectors (partial or total) are generic.

In Section~\ref{S-FFT12} we will also use singular properties of on-shell (dual) Bethe vectors
\begin{gather}\label{hwC}
\mathbb{C}_{a,b}(\blac;\bmuc) T_{ij}[0]=0,\qquad T_{ji}[0] \mathbb{B}_{a,b}(\blac;\bmuc)=0, \qquad i<j.
\end{gather}
Here $\mathbb{C}_{a,b}(\blac;\bmuc)$ and $\mathbb{B}_{a,b}(\blac;\bmuc)$ are on-shell Bethe vectors.
This property was found in \cite{MuhTV06} for ${\rm GL}(N)$-invariant models. In the ${\rm GL}(3)$ case it also follows from
the explicit formulas of the action of the operators $T_{ij}(z)$ onto Bethe vectors \cite{BelPRS12c}.

\section{Main results\label{S-MR}}

\begin{thm}\label{thm-anti}
Let $\mathbb{C}_{a',b'}(\blac;\bmuc)$ and $\mathbb{B}_{a,b}(\blab;\bmub)$ be total on-shell vectors such that
$\{\blac,\bmuc\}\ne\{\blab,\bmub\}$ $($that is, these on-shell vectors have different eigenvalues$)$. Then
\begin{gather}\label{FF-Tee-1}
\mathbb{C}_{a',b'}(\blac;\bmuc) T^{(1)}_{ij}[0] \mathbb{B}_{a,b}(\blab;\bmub)=
\left(\frac{\ell_1(\blac)\ell_3(\bmub)}{\ell_1(\blab)\ell_3(\bmuc)}-1\right){\mathfrak{F}}_{a,b}^{(i,j)}(\blac,\bmuc;\blab,\bmub),
\end{gather}
where ${\mathfrak{F}}_{a,b}^{(i,j)}$ is the universal form factor of the total operator $T_{ij}(z)$
and $a'=a+\delta_{i1}-\delta_{j1}$, $b'=b+\delta_{j3}-\delta_{i3}$.
\end{thm}

\begin{thm}\label{thm-diag}
Let $\mathbb{B}_{a,b}(\bla;\bmu)$ be  a total on-shell vector and $\mathbb{C}_{a,b}(\bla;\bmu)$ its dual
on-shell vector. Let $\mathbb{C}_{a,b}(\bla(\bar\kappa);\bmu(\bar\kappa))$ be a deformation of $\mathbb{C}_{a,b}(\bla;\bmu)$
such that the parameters $\bla(\bar\kappa)$ and $\bmu(\bar\kappa)$ satisfy twisted Bethe equations \eqref{ATEigenS-1}, and
$\bla(\bar\kappa)=\bla$, $\bmu(\bar\kappa)=\bmu$ at $\bar\kappa=1$. Then
\begin{gather}
\mathbb{C}_{a,b}(\bla;\bmu) T^{(1)}_{ii}[0] \mathbb{B}_{a,b}(\bla;\bmu)  = \!
\left(\!\delta_{i,1}\ell_1[0]+\delta_{i,3}\ell_3[0]+\frac{d}{d\kappa_i}
\log\frac{\ell_1\bigl(\bla(\bar\kappa)\bigr)}{\ell_3\bigl(\bmu(\bar\kappa)\bigr)}
\Bigr|_{\bar\kappa=1}\!\right)\!
 \|\mathbb{B}_{a,b}(\bla;\bmu)\|^2.\!\!\!\!\label{FF-Tee-2}
\end{gather}
\end{thm}

The proofs of these theorems will be given in the next
sections.

If the partial monodromy matrix $T^{(1)}(u)$ has the structure \eqref{mat-T12},
then the functions $\ell_k(u)$ actually depend also on the number $m$:
\begin{gather*}
\ell_k(u)=\prod_{n=1}^m\ell_k(u|n),\qquad k=1,3,
\end{gather*}
where  $\ell_k(u|n)$ are the local ratios
\begin{gather}\label{local-ratios}
 \ell_1(u|n)=\frac{\lambda_1(u|n)}{\lambda_2(u|n)}, \qquad  \ell_3(u|n)=\frac{\lambda_3(u|n)}{\lambda_2(u|n)}.
\end{gather}
In \eqref{local-ratios}, we introduced the vacuum eigenvalues of local $L$-operators $L_n(u)$
 \begin{gather*}
  (L_n(u) )_{ii}|0\rangle=\lambda_i(u|n)|0\rangle,\quad i=1,2,3.
 \end{gather*}

Using \eqref{part-zm} and Theorems~\ref{thm-anti},~\ref{thm-diag} we can f\/ind form factors of the local operators $ (L_m[0] )_{ij}$,
$i,j=1,2,3$. Namely, one has simply to consider the dif\/ference of two~$T^{(1)}(u)$ based on~$m$ and~$m-1$ respectively.

If $\{\blac,\bmuc\}\ne\{\blab,\bmub\}$, then we have
\begin{gather*}
 \mathbb{C}_{a',b'}(\blac;\bmuc) \bigl(L_m[0]\bigr)_{ij} \mathbb{B}_{a,b}(\blab;\bmub)
\nonumber\\
\qquad{} =
\left(\frac{\ell_1(\blac|m)\ell_3(\bmub|m)}{\ell_1(\blab|m)\ell_3(\bmuc|m)}-1\right) \left(\prod_{n=1}^{m-1}\frac{\ell_1(\blac|n)\ell_3(\bmub|n)}{\ell_1(\blab|n)\ell_3(\bmuc|n)}\right)
{\mathfrak{F}}_{a,b}^{(i,j)}(\blac,\bmuc;\blab,\bmub).
\end{gather*}

If $\{\blac,\bmuc\}=\{\blab,\bmub\}=\{\bla,\bmu\}$, then
\begin{gather*}
\mathbb{C}_{a,b}(\bla;\bmu) \bigl(L_m[0]\bigr)_{ij} \mathbb{B}_{a,b}(\bla;\bmu)=
\frac{d}{d\kappa_i} \log\frac{\ell_1\bigl(\bla(\bar\kappa)|m\bigr)}{\ell_3\bigl(\bmu(\bar\kappa)|m\bigr)}
\Bigr|_{\bar\kappa=1} \|\mathbb{B}_{a,b}(\bla;\bmu)\|^2.
\end{gather*}

If we deal with a continuum model, then form factors of local operators can be  found directly from~\eqref{FF-Tee-1},~\eqref{FF-Tee-2}.
In this case the integer number $m$ turns into a continuous variable $x$. This  parameter enters only the functions~$\ell_k$, and taking
the $x$-derivative of~\eqref{FF-Tee-1},~\eqref{FF-Tee-2} we f\/ind form factors of local operators in the point~$x$.

Thus, we obtain  form factors of local operators in the generalized model without use of a~specif\/ic
representation of the algebra \eqref{RTT}. In fact, it means that we have a solution of the quantum inverse scattering problem in
the weak sense. We cannot express the local operators in terms of the monodromy matrix entries as it was done in~\cite{KitMT99,MaiT00}, but we
can f\/ind all their matrix elements in the basis of the transfer matrix eigenstates. Furthermore, we have determinant formulas for all these
matrix elements~\cite{BelPRS12b,BelPRS13a,PakRS14c,PakRS14b,PakRS15a}.

{\bf $\boldsymbol{{\rm GL}(N)}$ generalisation.} We would like to mention that Theorems~\ref{thm-anti} and \ref{thm-diag} admit a direct generalization to ${\rm GL}(N)$-invariant models with $N>3$.
Indeed, Bethe vectors (and dual ones) of~${\rm GL}(N)$ models depend on $N-1$ sets of  parameters $\bar t^{j} = \{ t^{j}_1,t^{j}_2,\dots ,t^{j}_{a_j}\}$,
$j=1,2,\dots ,N-1$ and $N-1$ integers $a_j$ that correspond to the cardinalities of each set:
\begin{gather*}
\mathbb{B}_{\bar a}(\bar t) = \mathbb{B}_{a_1,a_2,\dots ,a_{N-1}}\big(\bar t^{1},\bar t^{2},\dots ,\bar t^{N-1}\big),\qquad
\mathbb{C}_{\bar a}(\bar t) = \mathbb{C}_{a_1,a_2,\dots ,a_{N-1}}\big(\bar t^{1},\bar t^{2},\dots ,\bar t^{N-1}\big).
\end{gather*}
The action of the  diagonal entries $T^{(l)}_{ii}$ on the vacuum vectors is similar to~\eqref{eigen}
\begin{gather*}
T_{ii}^{(l)}(t)|0\rangle^{(l)}= \lambda_{i}^{(l)}(t)|0\rangle^{(l)}, \qquad l=1,2,\quad i=1,\dots,N,
\end{gather*}
and we can introduce
\begin{gather*}
\alpha_i(t)=\frac{\lambda_{i}^{(1)}(t)}{\lambda_{i+1}^{(1)}(t)},\qquad i=1,\dots,N-1.
\end{gather*}
Note that in the case $N=3$ we have $\alpha_1(t)=\ell_1(t)$, while $\alpha_2(t)=\ell^{-1}_3(t)$.

\begin{Conj}\label{FF-Tee-1-GLN}
Form factors of the partial zero modes $T^{(1)}_{ij}[0]$ in ${\rm GL}(N)$-invariant models are given by
\begin{gather*}
\mathbb{C}_{\bar b}(\bar s) T^{(1)}_{ij}[0] \mathbb{B}_{\bar a}(\bar t)  =
\left(\prod_{k=1}^{N-1}\frac{\alpha_k(\bar s^{k})}{\alpha_k(\bar t^{k})}-1\right){\mathfrak{F}}_{\bar a}^{(i,j)}(\bar s;\bar t),
\qquad \mbox{for}\quad \bar s \neq\bar t
\\
\mathbb{C}_{\bar a}(\bar t) T^{(1)}_{ii}[0] \mathbb{B}_{\bar a}(\bar t)  =
\left( \lambda^{(1)}_{i}[0]+\sum_{k=1}^{N-1}\frac{d}{d\kappa_i}
\log\alpha_k\big(\bar t^{k}(\bar\kappa)\big)
\Bigr|_{\bar\kappa=1}\right)\|\mathbb{B}_{\bar a}(\bar t)\|^2.
\end{gather*}
where ${\mathfrak{F}}_{\bar a}^{(i,j)}$ is the universal $(z$-independent$)$ form factor of the total operator $T_{ij}(z)$
 and  we extended the convention~\eqref{SH-prod}  to the functions $\alpha_k$. $\bar t(\bar\kappa)$ is a set of $\kappa$-twisted on-shell Bethe parameters,
coinciding with $\bar t$ when $\bar\kappa=1$.
\end{Conj}

 This conjecture generalizes Theorems~\ref{thm-anti} and \ref{thm-diag}, proved for $N=2$ and $N=3$.

Remark that since $\mathbb{C}_{\bar b}(\bar s) T_{ij}[0] \mathbb{B}_{\bar a}(\bar t)=0$ when $\bar s\neq \bar t$, the above conjecture and theorems provide also the form factors for $T^{(2)}_{ij}[0]$.

\looseness=-1
It is worth mentioning that in the cases $N=2$ and $N=3$, compact determinant representations for the universal form factors are known. In contrast,
 in the case $N>3$ such representations are missing up to now. Nevertheless, if Conjecture~\ref{FF-Tee-1-GLN}
is valid in the~${\rm GL}(N)$ case, then it gives explicit dependence on the lattice site $m$ of the partial zero modes form factors.

\section{Form factors of diagonal operators}\label{S-FFDO}

We begin our consideration with the form factors of the diagonal partial zero modes $T^{(1)}_{ii}[0]$. It is convenient to
construct a special generating functional for these form factors \cite{IzeK84}.
Consider an operator
\begin{gather*}
Q_{\bar\beta}=\sum_{i=1}^3 \beta_i T^{(1)}_{ii}[0],
\end{gather*}
where $\beta_i$ are some complex numbers. The generating functional is
\begin{gather}\label{Gen-fun}
M_{a,b}^{(\bar\kappa)}= \mathbb{C}^{(\bar\kappa)}_{a,b}   (\blac;\bmuc) e^{Q_{\bar\beta}} \mathbb{B}_{a,b}(\blab;\bmub).
\end{gather}
Here $\mathbb{B}_{a,b}(\blab;\bmub)$ is an on-shell Bethe vector, $ \mathbb{C}^{(\bar\kappa)}_{a,b}(\blac;\bmuc)$ is
a dual  twisted on-shell Bethe vector with the twist parameters $\kappa_i=e^{\beta_i}$.

\begin{lem}\label{lemma-twist}
Let $M_{a,b}^{(\bar\kappa)}$ is defined as in~\eqref{Gen-fun}. Then
\begin{gather}\label{result}
M_{a,b}^{(\bar\kappa)}= e^{\beta_1\ell_1[0]+\beta_3\ell_3[0]}
\frac{\ell_1(\blac)\ell_3(\bmub)}{\ell_1(\blab)\ell_3(\bmuc)} \mathcal{S}^{(\bar\kappa)}_{a,b},
\end{gather}
where $\mathcal{S}^{(\bar\kappa)}_{a,b}$ is the scalar product of the twisted and the usual on-shell Bethe vectors~\eqref{Resh-SP-eig}.
\end{lem}

It is worth mentioning that an analog of~\eqref{result} for ${\rm GL}(2)$-based models was obtained in~\cite{KitKMST07}.
We will give a proof of  Lemma~\ref{lemma-twist}  in Section~\ref{S-PL}. Now we show how equation~\eqref{result} implies some statements
of Theorems~\ref{thm-anti}  and~\ref{thm-diag}.

Dif\/ferentiating   \eqref{Gen-fun}  over $\kappa_i$ at $\bar\kappa=1$ we obtain
\begin{gather}\label{Der-lhs}
\frac{d}{d\kappa_i}M_{a,b}^{(\bar\kappa)}\Bigr|_{\bar\kappa=1}=
\frac{d}{d\kappa_i}\mathcal{S}^{(\bar\kappa)}_{a,b}\Bigr|_{\bar\kappa=1}
+ \mathbb{C}^{(\bar\kappa)}_{a,b}(\blac;\bmuc)\Bigr|_{\bar\kappa=1}T^{(1)}_{ii}[0]\mathbb{B}_{a,b}(\blab;\bmub).
\end{gather}
Pay attention that the dual vector $\mathbb{C}_{a,b}(\blac;\bmuc)=\mathbb{C}^{(\bar\kappa)}_{a,b}(\blac;\bmuc)\Bigr|_{\bar\kappa=1}$ is an on-shell vector. Thus, the second
term in the r.h.s.\ of~\eqref{Der-lhs} is a form factor of the partial zero mode $T^{(1)}_{ii}[0]$.

On the other hand, dif\/ferentiating the r.h.s.\ of~\eqref{result}  over $\kappa_i$ at $\bar\kappa=1$ we f\/ind
\begin{gather}
\frac{d}{d\kappa_i}e^{\beta_1\ell_1[0]+\beta_3\ell_3[0]}
\frac{\ell_1(\blac)\ell_3(\bmub)}{\ell_1(\blab)\ell_3(\bmuc)} \mathcal{S}^{(\bar\kappa)}_{a,b}
\Bigr|_{\bar\kappa=1}=\frac{\ell_1(\blac)\ell_3(\bmub)}{\ell_1(\blab)\ell_3(\bmuc)}\Bigr|_{\bar\kappa=1}
\frac{d}{d\kappa_i}\mathcal{S}^{(\bar\kappa)}_{a,b}\Bigr|_{\bar\kappa=1}\nonumber\\
\qquad{}
+\delta_{\mathbb{C},\mathbb{B}}\left(\delta_{i,1}\ell_1 [0]+\delta_{i,3}\ell_3 [0]+\frac{d}{d\kappa_i}\log\frac{\ell_1(\bla)}{\ell_3(\bmu)}
\Bigr|_{\bar\kappa=1}\right) \|\mathbb{B}_{a,b}(\bla;\bmu)\|^2,\label{Der-rhs}
\end{gather}
where
\begin{gather*}
\delta_{\mathbb{C},\mathbb{B}}= \begin{cases}
1,&\text{if} \ \ \{\blac,\bmuc\}\bigr|_{\bar\kappa=1}=\{\blab,\bmub\},\vspace{1mm}\\
0,&\text{if} \ \  \{\blac,\bmuc\}\bigr|_{\bar\kappa=1}\ne\{\blab,\bmub\}.
\end{cases}
\end{gather*}
Deriving this formula we used the orthogonality of on-shell Bethe vectors depending on dif\/ferent Bethe parameters. Comparing equations~\eqref{Der-lhs} and~\eqref{Der-rhs} we immediately arrive at the statement of Theorem~\ref{thm-diag}. If
$\{\blac,\bmuc\}\bigr|_{\bar\kappa=1}\ne\{\blab,\bmub\}$, then we obtain
\begin{gather*}
\mathbb{C}_{a,b}(\blac;\bmuc)T^{(1)}_{ii}[0]\mathbb{B}_{a,b}(\blab;\bmub)=
\left(\frac{\ell_1(\blac)\ell_3(\bmub)}{\ell_1(\blab)\ell_3(\bmuc)}-1\right)\Bigr|_{\bar\kappa=1}
\frac{d}{d\kappa_i}\mathcal{S}^{(\bar\kappa)}_{a,b}\Bigr|_{\bar\kappa=1}.
\end{gather*}
It was proved in~\cite{BelPRS13a} that the $\kappa_i$-derivative of the scalar product $\mathcal{S}^{(\bar\kappa)}_{a,b}$
at $\bar\kappa=1$ is equal to the universal form factor of the operator $T_{ii}(z)$. Thus, the statement of Theorem~\ref{thm-anti}
is proved for partial zero modes~$T^{(1)}_{ii}[0]$.

\subsection{Proof of Lemma~\ref{lemma-twist}}\label{S-PL}

The proof of Lemma~\ref{lemma-twist} is  lengthy but straightforward.
 Let us f\/irst sketch the general strategy before going into details.
Knowing the action of the operator $e^{Q_{\bar\beta}}$ on
the partial Bethe vectors we f\/ind its action on the total Bethe vectors. Then we can can calculate the matrix
ele\-ment~$M_{a,b}^{(\bar\kappa)}$ in terms of scalar products of partial Bethe vectors, for which we use
equation \eqref{Resh-SP}. The resulting formula becomes rather cumbersome. In particular, it contains a sum over partitions of every set of the
original Bethe parameters into four subsets. Therefore in this section we use standard arabic indices in order to label these subsets.
New subsets of the Bethe parameters can be easily recombined into new sets of variables, and after
this the proof reduces to the use of identity~\eqref{Resh-SP-eig2} and equation~\eqref{Resh-SP-eig}.

Let us now give the details. Using \eqref{actBVd} and~\eqref{BV-BV-1}  we f\/ind the action of~$e^{Q_{\bar\beta}}$ on the total on-shell Bethe vector
\begin{gather*}
e^{Q_{\bar\beta}}\mathbb{B}_{a,b}(\blab;\bmub)=\sum  e^{\beta_1(\ell_1[0]-a_{\so})+\beta_2 (a_{\so}-b_{\so})+\beta_3(\ell_3[0]+ b_{\so})}
\frac{\ell_{3}(\bmub_{\st})}{\ell_{1}(\blab_{\so})} \\
\hphantom{e^{Q_{\bar\beta}}\mathbb{B}_{a,b}(\blab;\bmub)=}{}
\times f(\blab_{\so},\blab_{\st})f(\bmub_{\st},\bmub_{\so})
f(\bmub_{\so},\blab_{\so})
\mathbb{B}_{a_{\so},b_{\so}}^{(1)}(\blab_{\so};\bmub_{\so}) \mathbb{B}_{a_{\st},b_{\st}}^{(2)}(\blab_{\st};\bmub_{\st}).
\end{gather*}
Multiplying this equation from the left by $\mathbb{C}^{(\bar\kappa)}_{a,b}(\blac;\bmuc)$ and using~\eqref{CV-CV-eig}
we obtain
\begin{gather}
M_{a,b}^{(\bar\kappa)}=\sum
e^{\beta_1(\ell_1[0]-a_{\so})+\beta_2 a_{\so}+\beta_3\ell_3[0]}
\frac{\ell_1(\blac_{\st})\ell_3(\bmub_{\st})}{\ell_1(\blab_{\so})\ell_3(\bmuc_{\so})}
\nonumber\\
\hphantom{M_{a,b}^{(\bar\kappa)}=}{}
\times f(\blac_{\so},\blac_{\st})f(\blab_{\so},\blab_{\st})f(\bmuc_{\st},\bmuc_{\so})f(\bmub_{\st},\bmub_{\so})
f(\bmub_{\so},\blab_{\so})f(\bmuc_{\so},\blac_{\so})\nonumber\\
\hphantom{M_{a,b}^{(\bar\kappa)}=}{}
\times \mathbb{C}_{a_{\so},b_{\so}}^{(1)}(\blac_{\so};\bmuc_{\so})\mathbb{B}_{a_{\so},b_{\so}}^{(1)}(\blab_{\so};\bmub_{\so}) \cdot \mathbb{C}_{a_{\st},b_{\st}}^{(2)}(\blac_{\st};\bmuc_{\st})\mathbb{B}_{a_{\st},b_{\st}}^{(2)}(\blab_{\st};\bmub_{\st}).\label{Mat-el}
\end{gather}
Thus, \looseness=-1 we have obtained the expression for
$M_{a,b}^{(\bar\kappa)}$ in terms of scalar products of partial Bethe
vectors. Note that in spite of the sets~$\blab$ and~$\bmub$ satisfy  the Bethe equations~\eqref{AEigenS-1}, and
the sets~$\blac$ and~$\bmuc$ satisfy  the twisted Bethe equations~\eqref{ATEigenS-1}, the partial  Bethe vectors in~\eqref{Mat-el} are
not (twisted) on-shell vectors.
In other words we deal with the scalar products of generic Bethe vectors
in~\eqref{Mat-el}. Therefore, we do not write the additional superscript $(\bar\kappa)$ for the dual vectors and we should use \eqref{Resh-SP} for the calculation of their scalar products. Hereby, for the scalar product of the
vectors $\mathbb{C}^{(1)}$ and $\mathbb{B}^{(1)}$ we should replace in~\eqref{Resh-SP} the functions $r_k$ by $\ell_k$, while
for the scalar product of the
vectors~$\mathbb{C}^{(2)}$ and~$\mathbb{B}^{(2)}$ we should replace in~\eqref{Resh-SP} the functions~$r_k$ by~$r_k\ell^{-1}_k$.

The use of~\eqref{Resh-SP}  introduces new partitions of the subsets of Bethe parameters, so that, as mentioned above (see Section~\ref{SS-Not}), we use now arabic numbers to label the numerous subsubsets.
Thus, we have for the f\/irst scalar product
 \begin{gather*}
\mathbb{C}^{(1)}_{a,b}(\blac_{\so};\bmuc_{\so})\mathbb{B}^{(1)}_{a,b}(\blab_{\so};\bmub_{\so})=\sum \ell_1(\blab_{1})\ell_1(\blac_{3})
 \ell_3(\bmuc_{3})\ell_3(\bmub_{1})
  f(\blac_{1},\blac_{3})  f(\blab_{3},\blab_{1})     \\
 \qquad{}
 \times f(\bmuc_{3},\bmuc_{1})   f(\bmub_{1},\bmub_{3})\frac{f(\bmuc_{1},\blac_{1})f(\bmub_{3},\blab_{3})}
 {f(\bmuc_{\so},\blac_{\so})f(\bmub_{\so},\blab_{\so})}
   Z_{a_{3},b_{1}}(\blac_{3};\blab_{3}|\bmuc_{1};\bmub_{1})
 Z_{a_{1},b_{3}}(\blab_{1};\blac_{1}|\bmub_{3};\bmuc_{3}).
 \end{gather*}
 The summation is taken with respect to the partitions
\begin{gather*}
\blacb_{\so}\Rightarrow\{\blacb_1,\blacb_3\},\qquad \bmucb_{\so}\Rightarrow\{\bmucb_1,\bmucb_3\}.
\end{gather*}
The cardinalities of the subsubsets are $a_n=\#\blacb_n$, $b_n=\#\bmucb_n$, $n=1,3$.

Similarly
 \begin{gather}
\mathbb{C}^{(2)}_{a,b}(\blac_{\st};\bmuc_{\st})\mathbb{B}^{(2)}_{a,b}(\blab_{\st};\bmub_{\st})=\sum
\frac{r_1(\blab_{2})r_1(\blac_{4}) r_3(\bmuc_{4})r_3(\bmub_{2})}{\ell_1(\blab_{2})\ell_1(\blac_{4}) \ell_3(\bmuc_{4})\ell_3(\bmub_{2})}
  f(\blac_{2},\blac_{4})  f(\blab_{4},\blab_{2})   \label{Resh-SP-s2} \\
 \qquad{}
 \times f(\bmuc_{4},\bmuc_{2})   f(\bmub_{2},\bmub_{4})\frac{f(\bmuc_{2},\blac_{2})f(\bmub_{4},\blab_{4})}
 {f(\bmuc_{\st},\blac_{\st})f(\bmub_{\st},\blab_{\st})}
   Z_{a_{4},b_{2}}(\blac_{4};\blab_{4}|\bmuc_{2};\bmub_{2})
 Z_{a_{2},b_{4}}(\blab_{2};\blac_{2}|\bmub_{4};\bmuc_{4}).\nonumber
 \end{gather}
Here the sum is taken over partitions
\begin{gather*}
\blacb_{\st}\Rightarrow\{\blacb_2,\blacb_4\},\qquad \bmucb_{\st}\Rightarrow\{\bmucb_2,\bmucb_4\}.
\end{gather*}
The cardinalities of the subsubsets are still denoted by $a_n=\#\blacb_n$,\ and $b_n=\#\bmucb_n$, $n=2,4$.

Now we should express the products of the functions $r_k$ in~\eqref{Resh-SP-s2} via
 the (twisted) Bethe equations  for the full sets $\{\blab,\bmub\}$ and $\{\blac,\bmuc\}$.
We have
\begin{gather*}
 r_1(\blab_2)=\frac{f(\blab_2,\blab_1)f(\blab_2,\blab_3)f(\blab_2,\blab_4)}
 {f(\blab_1,\blab_2)f(\blab_3,\blab_2)f(\blab_4,\blab_2)} f(\bmub,\blab_2),
 \\
r_3(\bmub_2)=\frac{f(\bmub_1,\bmub_2)f(\bmub_3,\bmub_2)f(\bmub_4,\bmub_2)}
{f(\bmub_2,\bmub_1)f(\bmub_2,\bmub_3)f(\bmub_2,\bmub_4)}f(\bmub_2,\blab),
\\
 r_1(\blac_4)=e^{a_4(\beta_2-\beta_1)}\frac{f(\blac_4,\blac_1)f(\blac_4,\blac_2)f(\blac_4,\blac_3)}
 {f(\blac_1,\blac_4)f(\blac_2,\blac_4)f(\blac_3,\blac_4)}
 f(\bmuc,\blac_4),
\\
r_3(\bmuc_4)=e^{b_4(\beta_2-\beta_3)}\frac{f(\bmuc_1,\bmuc_4)f(\bmuc_2,\bmuc_4)f(\bmuc_3,\bmuc_4)}
{f(\bmuc_4,\bmuc_1)f(\bmuc_4,\bmuc_2)f(\bmuc_4,\bmuc_3)}f(\bmuc_4,\blac).
\end{gather*}

All these expressions should be substituted into \eqref{Mat-el}. After simple but exhausting algebra  we obtain
\begin{gather}
M_{a,b}^{(\bar\kappa)}=\sum
e^{\beta_1\ell_1[0]+\beta_3\ell_3[0]+(\beta_2-\beta_1) (a-a_2)+(\beta_2-\beta_3)b_4}
\frac{\ell_1(\blac_{2})\ell_1(\blac_{3})\ell_3(\bmub_{1})\ell_3(\bmub_{4})}{\ell_1(\blab_{2})\ell_1(\blab_{3})\ell_3(\bmuc_{1})\ell_3(\bmuc_{4})} \nonumber\\
\hphantom{M_{a,b}^{(\bar\kappa)}=}{}
\times F^C_{uu} F^C_{vv} F^C_{vu}
 F^B_{uu} F^B_{vv} F^B_{vu}   \mathcal{Z}.\label{Mat-el2}
\end{gather}
Here the sum is taken over partitions of every set of Bethe parameters into four subsets
\begin{gather*}
\blacb \Rightarrow\{\blacb_1,\blacb_2,\blacb_3,\blacb_4\},\qquad
\bmucb\Rightarrow\{\bmucb_1,\bmucb_2,\bmucb_3,\bmucb_4\}.
\end{gather*}
We have $\#\blab_n=\#\blac_n=a_n$ and $\#\bmub_n=\#\bmuc_n=b_n$, $n=1,\dots ,4$, but the values~$a_n$ and~$b_n$ are free. Note that~$a_2$ and~$b_4$ explicitly appear as coef\/f\/icients in~\eqref{Mat-el2}, so that manipulations with~$\blacb_2$ and~$\bmucb_4$ should include these coef\/f\/icients.

The factor $\mathcal{Z}$ in~\eqref{Mat-el2} is the product of four highest coef\/f\/icients
\begin{gather*}
\mathcal{Z}  =  Z_{a_3,b_1}(\blac_3;\blab_3|\bmuc_1;\bmub_1)  Z_{a_1,b_3}(\blab_1;\blac_1|\bmub_3;\bmuc_3)
Z_{a_4,b_2}(\blac_4;\blab_4|\bmuc_2;\bmub_2)
  Z_{a_2,b_4}(\blab_2;\blac_2|\bmub_4;\bmuc_4).
\end{gather*}
The other factors in~\eqref{Mat-el2} denoted by~$F$ with dif\/ferent subscripts and superscripts are
products of $f$ functions:
\begin{gather*}
F^C_{uu}=f(\blac_4,\blac_1)
f(\blac_3,\blac_2)
f(\blac_4,\blac_2)
f(\blac_4,\blac_3)
f(\blac_1,\blac_2)
f(\blac_1,\blac_3),
\\
F^B_{uu}=f(\blab_1,\blab_4)
f(\blab_2,\blab_3)
f(\blab_2,\blab_1)
f(\blab_2,\blab_1)
f(\blab_3,\blab_4)
f(\blab_3,\blab_4),
\\
F^C_{vv}=f(\bmuc_1,\bmuc_4)
f(\bmuc_2,\bmuc_3)
f(\bmuc_2,\bmuc_1)
f(\bmuc_2,\bmuc_4)
f(\bmuc_3,\bmuc_1)
f(\bmuc_3,\bmuc_4),
\\
F^B_{vv}=f(\bmub_4,\bmub_1)
f(\bmub_3,\bmub_2)
f(\bmub_1,\bmub_3)
f(\bmub_4,\bmub_3)
f(\bmub_1,\bmub_2)
f(\bmub_4,\bmub_2),
\\
F^C_{vu}=f(\bmuc_1,\blac_4)f(\bmuc_4,\blac_4)f(\bmuc_1,\blac_1)f(\bmuc_4,\blac_1)
f(\bmuc_3,\blac_4)f(\bmuc_4,\blac_3),
\\
F^B_{vu}=f(\bmub_3,\blab_3)f(\bmub_2,\blab_2)f(\bmub_3,\blab_2)f(\bmub_2,\blab_3)
f(\bmub_1,\blab_2)f(\bmub_2,\blab_1).
\end{gather*}

It remains to combine the subsubsets into new groups:
\begin{alignat*}{3}
& \{\blacb_1,\blacb_4\} =\blacb_{\rm i},\qquad && \{\blacb_2,\blacb_3\} =\blacb_{\rm ii},& \\
& \{\bmucb_1,\bmucb_4\} =\bmucb_{\rm i},\qquad && \{\bmucb_2,\bmucb_3\} =\bmucb_{\rm ii}.&
\end{alignat*}
Then we recast \eqref{Mat-el2} as follows:
\begin{gather}
M_{a,b}^{(\bar\kappa)}=\sum_{\substack{\blacb\Rightarrow\{\blacb_{\rm i},\blacb_{\rm ii}\}\\
\bmucb\Rightarrow\{\bmucb_{\rm i},\bmucb_{\rm ii}\}}}
\frac{\ell_1(\blac_{\rm ii})\ell_3(\bmub_{\rm i})}{\ell_1(\blab_{\rm ii})\ell_3(\bmuc_{\rm i})}
f(\blac_{\rm i},\blac_{\rm ii})f(\blab_{\rm ii},\blab_{\rm i})f(\bmuc_{\rm ii},\bmuc_{\rm i})f(\bmub_{\rm i},\bmub_{\rm ii})\nonumber\\
\hphantom{M_{a,b}^{(\bar\kappa)}=}{}
\times f(\bmuc_{\rm i},\blac_{\rm i})f(\bmub_{\rm ii},\blab_{\rm ii}) G_1(\blac_{\rm i},\blab_{\rm i};\bmuc_{\rm ii},\bmub_{\rm ii})
 G_2(\blac_{\rm ii},\blab_{\rm ii};\bmuc_{\rm i},\bmub_{\rm i}),\label{Mat-el4}
\end{gather}
where factors $G_1$ and $G_2$ are given as sums over partitions
 \begin{gather}
 G_1(\blac_{\rm i},\blab_{\rm i};\bmuc_{\rm ii},\bmub_{\rm ii})=\sum_{\substack{\blacb_{\rm i}\Rightarrow\{\blacb_{1},\blacb_{4}\}\\
\bmucb_{\rm ii}\Rightarrow\{\bmucb_{2},\bmucb_{3}\}}}   f(\blac_{4},\blac_{1})  f(\blab_{1},\blab_{4})  f(\bmuc_{2},\bmuc_{3})
   f(\bmub_{3},\bmub_{2})
  \nonumber\\
\hphantom{G_1(\blac_{\rm i},\blab_{\rm i};\bmuc_{\rm ii},\bmub_{\rm ii})=}{}
 \times f(\bmuc_{3},\blac_{4})f(\bmub_{2},\blab_{1})  Z_{a_{4},b_{2}}(\blac_{4};\blab_{4}|\bmuc_{2};\bmub_{2})
 Z_{a_{1},b_{3}}(\blab_{1};\blac_{1}|\bmub_{3};\bmuc_{3}),\label{G1}
 \end{gather}
and
 \begin{gather}
  G_2(\blac_{\rm ii},\blab_{\rm ii};\bmuc_{\rm i},\bmub_{\rm i})=\sum_{\substack{\blacb_{\rm ii}\Rightarrow\{\blacb_{2},\blacb_{3}\}\\
\bmucb_{\rm i}\Rightarrow\{\bmucb_{1},\bmucb_{4}\}}}  e^{\beta_1\ell_1[0]+\beta_3\ell_3[0]+(\beta_2-\beta_1) (a-a_2)+(\beta_2-\beta_3)b_4}
 \nonumber\\
\hphantom{G_2(\blac_{\rm ii},\blab_{\rm ii};\bmuc_{\rm i},\bmub_{\rm i})=}{}
 \times  f(\blac_{3},\blac_{2})
f(\blab_{2},\blab_{3}) f(\bmuc_{1},\bmuc_{4})
   f(\bmub_{4},\bmub_{1})f(\bmuc_{4},\blac_{3})f(\bmub_{1},\blab_{2})
   \nonumber\\
\hphantom{G_2(\blac_{\rm ii},\blab_{\rm ii};\bmuc_{\rm i},\bmub_{\rm i})=}{}
 \times Z_{a_{3},b_{1}}(\blac_{3};\blab_{3}|\bmuc_{1};\bmub_{1})
 Z_{a_{2},b_{4}}(\blab_{2};\blac_{2}|\bmub_{4};\bmuc_{4}).\label{G2}
 \end{gather}
It is not dif\/f\/icult to see that the sum over partitions in \eqref{G1} coincides with the sum in \eqref{Resh-SP-eig2}
up to relabeling of the subsets: $\blacb_1\to\blacb_{\so}$,  $\blacb_4\to\blacb_{\st}$,  $\bmucb_2\to\bmucb_{\so}$, and  $\bmucb_3\to\bmucb_{\st}$.
Thus,  we conclude that $G_1=0$ unless
$\blacb_{\rm i}=\varnothing$ and $\bmucb_{\rm ii}=\varnothing$. Hence, $\blacb_{\rm ii}=\blacb$,
$\bmucb_{\rm i}=\bmucb$, and $a_1=a_4=0$, $b_2=b_3=0$. Then due to~\eqref{Exception} $G_1=1$.

Looking now at \eqref{G2} and comparing it with~\eqref{Resh-SP-eig} we see that they coincide up to the common prefactor
$e^{\beta_1\ell_1[0]+\beta_3\ell_3[0]}$ and
relabeling of the subsets: $\blacb_2\to\blacb_{\so}$,  $\blacb_3\to\blacb_{\st}$,  $\bmucb_1\to\bmucb_{\so}$, and  $\bmucb_4\to\bmucb_{\st}$.
Hence,
\begin{gather*}
G_2= e^{\beta_1\ell_1[0]+\beta_3\ell_3[0]}\mathcal{S}^{(\bar\kappa)}_{a,b}.
\end{gather*}
Substituting this into \eqref{Mat-el4} and setting there $\blacb_{\rm i}=\varnothing$ and $\bmucb_{\rm ii}=\varnothing$ we immediately
arrive at~\eqref{result}.

\section{Form factor for  of\/f-diagonal partial zero modes}\label{S-FFT12}

Now we study the form factors of of\/f-diagonal partial zero modes.
 We apply a strategy similar to the one used in \cite{PakRS14c}, using the commutation relations of the zero modes and properties of the type~\eqref{actBVa},~\eqref{actBVd}.

 First, we note that the $RTT$-relation \eqref{RTT} implies in particular for the (partial) zero modes
\begin{alignat*}{3}\label{7comrel}
&[T_{ii}[0] , T_{ji}[0]] = T_{ji}[0], \qquad && [T^{(l)}_{ii}[0] , T^{(l)}_{ji}[0]] = T^{(l)}_{ji}[0] ,\quad i\neq j ,& \\
&[T_{ij}[0] , T_{ii}[0]] = T_{ij}[0], \qquad && [T^{(l)}_{ij}[0] , T^{(l)}_{ii}[0]] = T^{(l)}_{ij}[0] ,\quad i\neq j ,& \\
&[T_{ij}[0] , T_{ki}[0]] = T_{kj}[0], \qquad && [T^{(l)}_{ij}[0] , T^{(l)}_{ki}[0]] = T^{(l)}_{kj}[0] ,\quad i\neq j\neq k ,\qquad l=1,2.&
\end{alignat*}
Now, using $T_{kl}[0]=T^{(1)}_{kl}[0]+T^{(2)}_{kl}[0]$ and $[T^{(1)}_{ij}[0] , T^{(2)}_{kl}[0]]=0$ we conclude
\begin{gather}
[T^{(1)}_{ii}[0] , T_{ji}[0]] = T^{(l)}_{ji}[0] ,\quad i\neq j ,\nonumber\\
[T_{ij}[0] , T^{(1)}_{ii}[0]] = T^{(1)}_{ij}[0] ,\quad i\neq j ,\nonumber\\
[T_{ij}[0] , T^{(1)}_{ki}[0]] = T^{(1)}_{kj}[0] ,\quad i\neq j\neq k ,\label{full1}
\end{gather}
which are the central relations that we will use for our calculations.

As a notation, we will note the form factor of $T^{(1)}_{ij}[0]$ as
\begin{gather*}
M_{a,b}^{(i,j)}(\blac,\bmuc ; \blab,\bmub)=\mathbb{C}_{a',b'}(\blac;\bmuc) T^{(1)}_{ij}[0] \mathbb{B}_{a,b}(\blab;\bmub),  \qquad i,j=1,2,3,
\end{gather*}
with $a'=a+\delta_{i1}-\delta_{j1}$ and $b'=b+\delta_{j3}-\delta_{i3}$, and where both vectors are on-shell.

\subsection[Form factor of $\protect{T^{(1)}_{12}[0]}$]{Form factor of $\boldsymbol{T^{(1)}_{12}[0]}$}
We start with $M_{a,b}^{(1,2)}$, the form factor of $T^{(1)}_{12}[0]$. We make the calculation in details, the other ones following the same steps.
To get $M_{a,b}^{(1,2)}$, we start with the form factor $M_{a+1,b}^{(2,2)}$ and send one of the parameters in $\blab$ to inf\/inity (keeping all the other parameters f\/inite), using relations~\eqref{actBVa0}:
\begin{gather*}
\lim_{w\to\infty}\frac wc M_{a+1,b}^{(2,2)}(\blac,\bmuc ; \{\blab,w\},\bmub)
 = \lim_{w\to\infty}\frac wc\mathbb{C}_{a+1,b}(\blac;\bmuc) T^{(1)}_{22}[0] \mathbb{B}_{a+1,b}(\{\blab,w\};\bmub)
\nonumber\\
\hphantom{\lim_{w\to\infty}\frac wc M_{a+1,b}^{(2,2)}(\blac,\bmuc ; \{\blab,w\},\bmub)}{}
 = \mathbb{C}_{a+1,b}(\blac;\bmuc) T^{(1)}_{22}[0] T_{12}[0] \mathbb{B}_{a,b}(\blab;\bmub).
\end{gather*}
Now, since $\mathbb{C}_{a+1,b}(\blac;\bmuc)$ is on-shell with all parameters f\/inite, it is annihilated by $T_{12}[0]$  due to~\eqref{hwC}.
Thus, we can replace the product $T^{(1)}_{22}[0] T_{12}[0]$ by its commutator which in turn gives $T_{12}^{(1)}[0]$ through~\eqref{full1}.
It leads to
\begin{gather}\label{12M}
\lim_{w\to\infty}\frac wc M_{a+1,b}^{(2,2)}(\blac,\bmuc ; \{\blab,w\},\bmub)=
M_{a,b}^{(1,2)}(\blac,\bmuc ; \blab,\bmub).
\end{gather}
It remains to compute the limit of the explicit expression \eqref{FF-Tee-1} for $M_{a+1,b}^{(2,2)}$, that was proved   in the previous section. It is obvious that $\lim\limits_{w\to\infty} \ell_k(w)=1$, $k=1,3$, and it has been shown in~\cite{PakRS14c} that\footnote{Strictly speaking the proof of~\cite{PakRS14b}
was done  for the full form factor, but it extends straightforwardly to the universal form factor.}
\begin{gather}\label{limUF}
\lim_{w\to\infty}\frac wc {\mathfrak{F}}_{a+1,b}^{(2,2)}(\blac,\bmuc;\{\blab,w\},\bmub)={\mathfrak{F}}_{a,b}^{(1,2)}(\blac,\bmuc;\blab,\bmub).
\end{gather}
Hence, we get expression \eqref{FF-Tee-1} for $M_{a,b}^{(1,2)}$.

\subsection{Other form factors}\label{S-OFF}

As already mentioned, the calculation for other form factors follows the same steps, so we just sketch the proofs.

{\bf Form factor of $\boldsymbol{T^{(1)}_{23}[0]}$.} To get $M_{a,b}^{(2,3)}$,
we start with $M_{a,b+1}^{(2,2)}(\blac,\bmuc ; \blab ,\{\bmub,w\})$ and take the  limit ${w\to\infty}$.  It makes appear $T_{23}[0]$ that annihilates $\mathbb{C}_{a,b+1}(\blac;\bmuc)$. It leads to
\begin{gather}\label{23M}
\lim_{w\to\infty}\frac wc M_{a,b+1}^{(2,2)}(\blac,\bmuc ; \blab,\{\bmub,w\})=
-M_{a,b}^{(2,3)}(\blac,\bmuc ; \blab,\bmub).
\end{gather}

{\bf Form factor of $\boldsymbol{T^{(1)}_{21}[0]}$.} To get $M_{a,b}^{(2,1)}$,
we start with $M_{a,b}^{(2,2)}(\{\blac,w\},\bmuc ; \blab,\bmub)$ and take the  limit ${w\to\infty}$.
It makes appear $T_{21}[0]$ on the left, and it annihilates $\mathbb{B}_{a,b}(\blac;\bmuc)$  due to~\eqref{hwC}.
Then, we can again replace the product $T_{21}[0]T^{(1)}_{22}[0]$ by its commutator, and
we obtain
\begin{gather}\label{21M}
\lim_{w\to\infty}\frac wc M_{a,b}^{(2,2)}(\{\blac,w\},\bmuc  ; \blab,\bmub)=
M_{a,b}^{(2,1)}(\blac,\bmuc ; \blab,\bmub).
\end{gather}

{\bf Form factor of $\boldsymbol{T^{(1)}_{32}[0]}$.} The form factor $M_{a,b}^{(3,2)}$ is obtained through the limit
\begin{gather}\label{32M}
\lim_{w\to\infty}\frac{w}{c} M_{a,b}^{(2,2)}(\blac,\{\bmuc,w\} ; \blab,\bmub)=
- M_{a,b}^{(3,2)}(\blac,\bmuc ; \blab,\bmub).
\end{gather}

{\bf Form factor of $\boldsymbol{T^{(1)}_{13}[0]}$.} 
To get $M_{a,b}^{(1,3)}$,
it is convenient to start with  the already known form factor
$M_{a+1,b}^{(2,3)}(\blac,\bmuc ; \{\blab,w\},\bmub)$. Taking the  limit $ w\to\infty$ we obtain $T_{12}[0]$ that annihilates $\mathbb{C}_{a+1,b}(\blac;\bmuc)$. Hence, due to the last equation \eqref{full1} we obtain
\begin{gather}\label{13M}
\lim_{w\to\infty}\frac{w}{c} M_{a+1,b}^{(2,3)}(\blac,\bmuc ; \{\blab,w\},\bmub)=
M_{a,b}^{(1,3)}(\blac,\bmuc ; \blab,\bmub).
\end{gather}

{\bf Form factor of $\boldsymbol{T^{(1)}_{31}[0]}$.} To get $M_{a,b}^{(3,1)}$,
we start with $M_{a,b}^{(3,2)}(\{\blac,w\},\bmuc ; \blab,\bmub)$ and take the  limit  $ w\to\infty$.
 We obtain
\begin{gather}\label{31M}
\lim_{w\to\infty}\frac{w}{c} M_{a,b}^{(3,2)}(\{\blac,w\}, \bmuc;\blab,\bmub)=
M_{a,b}^{(3,1)}(\blac,\bmuc ; \blab,\bmub).
\end{gather}

Thus, starting from one initial form factor $M_{a,b}^{(2,2)}$ we can obtain all other form factors $M_{a,b}^{(i,j)}$ in the special limits of the Bethe
parameters. This property is a direct consequence of the property of the form factors of the monodromy matrix entries $\mathcal{F}_{a,b}^{(i,j)}$~\cite{PakRS15a}.

{\bf Consistency with morphisms.}
It is worth mentioning that there exist also other relations between dif\/ferent form factors $M_{a,b}^{(i,j)}$ \cite{PakRS14b,PakRS15a}.
These relations appear due to morphisms of the algebra~\eqref{RTT}. It was shown in~\cite{BelPRS12c} that  the  mappings
\begin{gather}\label{def-psi}
\psi\colon \
T_{ij}(u)  \mapsto  T_{ji}(u),\qquad
\varphi\colon \  T_{ij}(u)  \mapsto  T_{4-j,4-i}(-u),
\end{gather}
def\/ine morphisms of the algebra \eqref{RTT}. Hereby, the mapping $\varphi$ is an isomorphism, while $\psi$ is an antimorphism.
Both mappings \eqref{def-psi}  exchange the partial monodromy matrices of the composite model \cite{PakRS15b}
\begin{gather*}
\psi\colon \  T^{(l)}_{ij}(u)   \mapsto  T^{(3-l)}_{ji}(u),\qquad
\varphi\colon \  T^{(l)}_{ij}(u)  \mapsto  T^{(3-l)}_{4-j,4-i}(-u).
\end{gather*}

Transformations \eqref{def-psi} induce relations between dif\/ferent form factors of the monodromy matrix entries $\mathcal{F}_{a,b}^{(i,j)}$
(see \cite{PakRS14c, PakRS14b} for details).
The latest, in their turn, give us relations between the universal form factors $\mathfrak{F}_{a,b}^{(i,j)}$
\begin{gather}
\mathfrak{F}_{a,b}^{(i,j)}(\blac,\bmuc;\blab,\bmub) =-\mathfrak{F}_{a',b'}^{(j,i)}(\blab,\bmub;\blac,\bmuc),\nonumber\\
\mathfrak{F}_{a,b}^{(i,j)}(\blac,\bmuc;\blab,\bmub) =
\mathfrak{F}_{b,a}^{(4-j,4-i)}(-\bmuc,-\blac;-\bmub,-\blab).\label{psiFF-u}
\end{gather}
Thus, actually it is enough to compute only four form factors of the partial zero modes~$T^{(1)}_{1j}[0]$ ($j=1,2,3$) and
$T^{(1)}_{22}[0]$. All other form factors can be obtained by the mappings described above. Using the explicit determinant
representations for the from factors one can check that they enjoy both the limiting procedures~\eqref{12M}, \eqref{23M}--\eqref{31M},
and the transformations~\eqref{psiFF-u}.

{\bf Generalisation to $\boldsymbol{{\rm GL}(N)}$.}
One can apply the same procedure to the ${\rm GL}(N)$ case. In particular, the commutation relations \eqref{full1} are still valid, and the limits of the type \eqref{actBVa}, \eqref{actBVd} and \eqref{limUF} have been proven for ${\rm GL}(N)$ in~\cite{PakRS14c}.
The singular vector properties of the type~\eqref{hwC}  were obtained for ${\rm GL}(N)$ in~\cite{MuhTV06}.
All that allows us to relate the dif\/ferent form factors in the same way we did for ${\rm GL}(3)$. It is easy to check that the Conjecture~\eqref{FF-Tee-1-GLN} is consistent with these relations.

\section{Conclusion}

In this paper we have studied form factors of the partial zero modes in a composite  generalized model with ${\rm GL}(3)$-invariant $R$-matrix. We have reduced
these form factors to the ones
of the monodromy matrix entries $T_{ij}(u)$ considered in our previous publications. As we have mentioned already, it means that
we have a solution of the inverse scattering problem in the weak sense. Apparently the same type of the formulas remains true for ${\rm GL}(N)$-invariant
models with $N>3$.

We are planning to apply these results to the study of form factors of local operators in the model of one-dimensional two-component Bose gas with $\delta$-function interaction \cite{Sath68, Yang67}. Several form factors in this model were studied already in the framework of the coordinate Bethe ansatz for some particular cases of Bethe vectors~\cite{PozOK12}. We are going to apply the algebraic Bethe ansatz techniques in order to compute the form factors in the general case.
This model possesses ${\rm GL}(3)$-invariant $R$-matrix~\eqref{R-mat} and can be considered in the framework of the scheme described
in the present paper. However the asymptotic expansion of the monodromy matrices~\eqref{zero-modes} in the case of Bose gas should be modif\/ied.
In its turn, this modif\/ication leads to a modif\/ication of the partial zero modes. Nevertheless  the derivation of determinant representations
for the form factors of the diagonal partial zero modes~$T^{(1)}_{ii}[0]$ does not change. A possibility to use this result for obtaining all other
form factors is not obvious, however they can be calculated by a straightforward method similar to the one that we used for the calculation
of $T^{(1)}_{ii}[0]$ in the present paper. This will be the subject of our further publication.

\subsection*{Acknowledgements}

The work of S.P.~was supported in part by RFBR-Ukraine grant 14-01-90405-ukr-a.
N.A.S.~was  supported by the Program of RAS  ``Nonlinear Dynamics in Mathematics and Physics'',
RFBR-14-01-00860-a, RFBR-13-01-12405-of\/i-m2.

\pdfbookmark[1]{References}{ref}
\LastPageEnding

\end{document}